\documentclass[fleqn,usenatbib]{mnras}

\usepackage{newtxtext,newtxmath}

\usepackage[T1]{fontenc}

\DeclareRobustCommand{\VAN}[3]{#2}
\let\VANthebibliography\thebibliography
\def\thebibliography{\DeclareRobustCommand{\VAN}[3]{##3}\VANthebibliography}

\usepackage{graphicx}	
\usepackage{amsmath}	
\usepackage{color}
\usepackage{soul}

\title[The 100\,pc \emph{Gaia} sample of unresolved WDMS binaries]{White dwarf-main sequence binaries from \emph{Gaia} EDR3: the unresolved 100\,pc volume-limited sample}

\author[Rebassa-Mansergas et
  al.]{A. Rebassa-Mansergas$^{1,2}$\thanks{E-mail:
    alberto.rebassa@upc.edu}, E. Solano$^{3,4}$, F. M. Jim\'enez-Esteban$^{3,4}$, S. Torres$^{1,2}$, C. Rodrigo$^{3,4}$, \newauthor A. Ferrer-Burjachs$^{1}$,  L. M. Calcaferro$^{5,6}$, L. G. Althaus$^{5,6}$, A. H. C\'orsico$^{5,6}$
   \\
$^{1}$ Departament de F\'{\i}sica, Universitat Polit\`{e}cnica de Catalunya, c/Esteve Terrades 5, 08860 Castelldefels, Spain\\
$^{2}$ Institut d'Estudis Espacials de Catalunya, Ed. Nexus-201, c/Gran Capit\`a 2-4, 08034 Barcelona, Spain\\
$^{3}$Departmento de Astrof\'{\i}sica, Centro de Astrobiolog\'{\i}a (CSIC-INTA), ESAC Campus, Camino Bajo del Castillo s/n,\\
E-28692 Villanueva de la Ca\~nada, Madrid, Spain\\
$^{4}$ Spanish Virtual Observatory, E-28692 Villanueva de la Ca\~nada, Madrid, Spain\\
$^{5}$ Grupo de Evoluci\'on Estelar y Pulsaciones, Facultad de Ciencias Astron\'omicas y Geof\'isicas, Universidad Nacional de La Plata,\\ Paseo del Bosque s/n, 1900, La Plata, Argentina\\
$^{6}$ Instituto de Astrof\'isica La Plata, CONICET-UNLP, Paseo del Bosque s/n, 1900, La Plata, Argentina
}

\date{Accepted XXX. Received YYY; in original form ZZZ}

\pubyear{2021}

\begin{document}
\label{firstpage}
\pagerange{\pageref{firstpage}--\pageref{lastpage}}
\maketitle

\begin{abstract}
We use  the data provided by  the \emph{Gaia} Early Data  Release 3 to
search  for  a  highly-complete volume-limited  sample  of  unresolved
binaries consisting  of a  white dwarf and  a main  sequence companion
(i.e. WDMS binaries)  within 100\,pc.  We select 112  objects based on
their location within the Hertzsprung-Russell diagram, of which 97 are
new identifications.  We fit their spectral energy distributions (SED)
with  a  two-body  fitting  algorithm  implemented  in  VOSA  (Virtual
Observatory  SED  Analyser)  to  derive  the  effective  temperatures,
luminosities and  radii (hence surface  gravities and masses)  of both
components.  The  stellar parameters  are compared  to those  from the
currently largest  catalogue of  close WDMS  binaries, from  the Sloan
Digital Sky Survey (SDSS).  We  find important differences between the
properties of  the \emph{Gaia} and  SDSS samples.  In  particular, the
\emph{Gaia}  sample contains  WDMS binaries  with considerably  cooler
white dwarfs and  main sequence companions (some expected  to be brown
dwarfs). The \emph{Gaia} sample also  shows an important population of
systems consisting  of cool and  extremely low-mass white  dwarfs, not
present in the  SDSS sample.  Finally, using a  Monte Carlo population
synthesis  code, we  find that  the volume-limited  sample of  systems
identified here seems to be highly complete ($\simeq80\pm9$ per cent),
however it only represents $\simeq$9  per cent of the total underlying
population.  The missing $\simeq$91 per cent includes systems in which
the  main sequence  companions  entirely dominate  the  SEDs. We  also
estimate  an upper  limit to  the total  space density  of close  WDMS
binaries of $\simeq(3.7\pm1.9)\times 10^{-4}$\,pc$^{-3}$.
\end{abstract}

\begin{keywords}
(stars:) white dwarfs -- (stars:) binaries (including multiple): close
  -- virtual observatory tools
\end{keywords}



\section{Introduction}

Binary and  multiple stellar systems  are common. The  binary fraction
among early-type  O and B stars  is over 70 per  cent \citep{Sana2014,
  Moe+distefano2017}; the probability of stars  like our Sun for being
in  binaries   is  $\simeq$50  per   cent  \citep{Duquennoy+Mayor1991,
  Raghavan2010} and  this value decreases  to $\simeq$30 per  cent for
the lowest mass main sequence  stars, the M dwarfs \citep{Siegler2005,
  Winters2019}. It  is therefore unquestionable that  binary stars are
an important ingredient in the study of stellar evolution.

If the binary  components are separated enough to  avoid mass transfer
episodes ($\ga$10  AU; \citealt{Farihi2010}), their  evolution follows
that of  single stars. In this  sense, the more massive  main sequence
star in the binary evolves  through the typical nuclear burning phases
at a faster pace  than its lower mass companion and, if  it has a mass
of $\leq$ 8--11  M$_{\odot}$ -- this is  true for over 95  per cent of
the   stars    --,   it   ends    its   life   as   a    white   dwarf
\citep{GarciaBerro1997}. These  white dwarf plus main  sequence (WDMS)
binaries  have orbital  separations that  are similar  to the  initial
ones, or  wider due to mass  loss of the white  dwarf progenitors that
results in the expansion of the orbits. Wide WDMS binaries can be used
to constrain  the age-metallicity relation in  the solar neighbourhood
\citep{Rebassa2016, Rebassa2021},  the initial-to-final  mass relation
\citep{Catalan2008, Zhao2012},  the age-activity-rotation  relation of
low-mass main sequence stars  \citep{Rebassa2013, Skinner2017} and the
binary star  separation for a  planetary system origin of  white dwarf
pollution \citep{Veras2018}, among other examples.

\begin{figure*}
    \centering
    \includegraphics[width=\columnwidth]{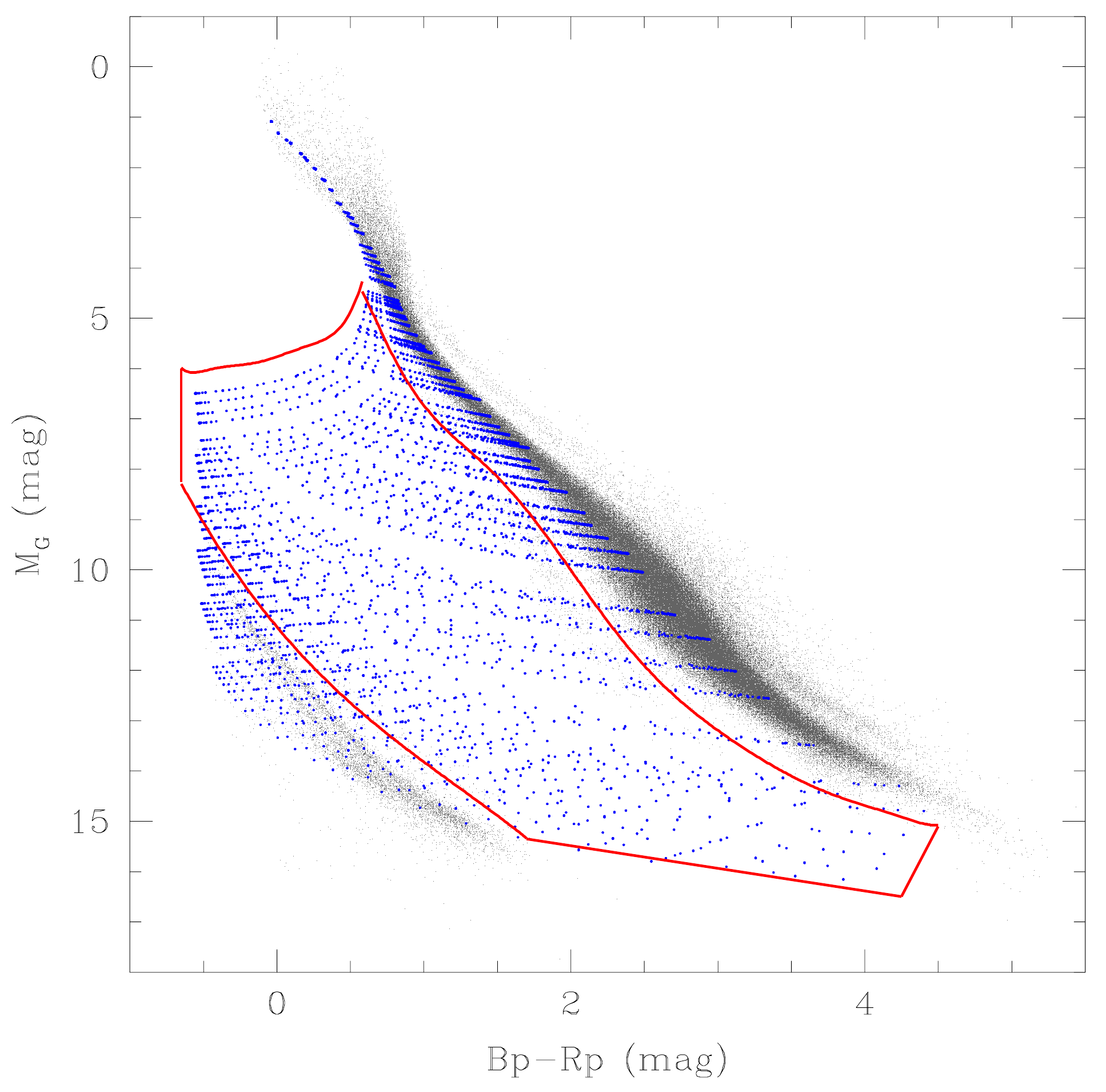}
    \includegraphics[width=\columnwidth]{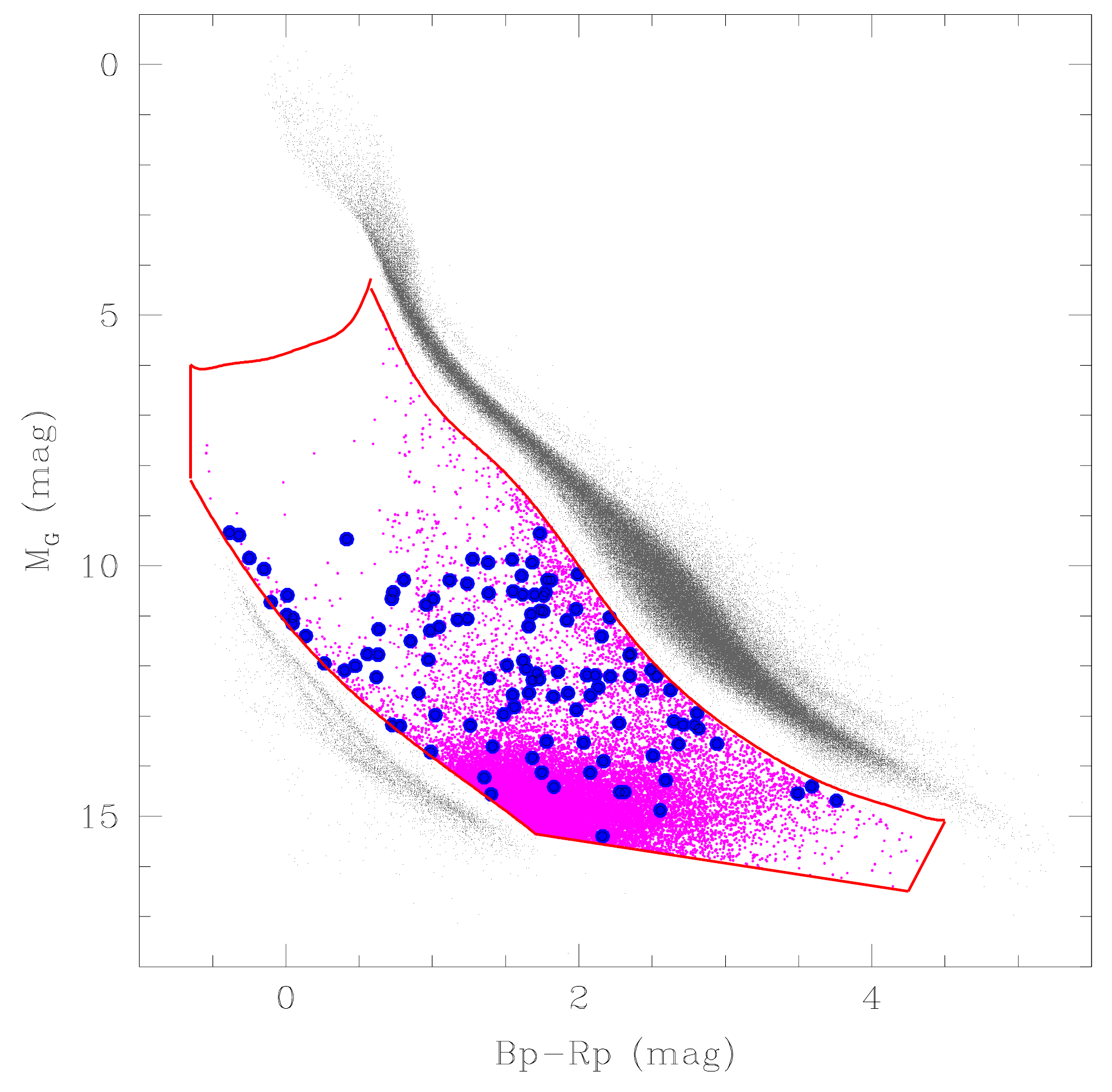}
    \caption{Left  panel: \emph{Gaia}  EDR3  main  sequence stars  and
      white  dwarfs within  100\,pc  (gray dots),  the synthetic  WDMS
      binaries (blue dots) and  the colour-magnitude selection of WDMS
      binaries (red solid lines).  The absolute magnitudes and colours
      of the synthetic WDMS binaries were obtained combining a grid of
      white dwarf  (with effective temperatures and  surface gravities
      ranging from  3\,000 to 100\,000\,K and  from 7 to 9.5  dex) and
      main  sequence star  (with effective  temperatures ranging  from
      2\,400  to  9\,700\,K) fluxes.   See  text  for details.   Right
      panel: The  same as the left  panel but showing the  WDMS binary
      candidates (magenta) that  result from applying the  cuts to the
      \emph{Gaia}  EDR3 data  base. The  final sample  after excluding
      contaminants is shown as blue solid dots.}
    \label{fig-HR}
\end{figure*} 

Conversely, for smaller initial main sequence binary separations, mass
transfer interactions ensue, which generally lead to a common envelope
phase    and    to    a    dramatic    shrinkage    of    the    orbit
\citep{Willems+Kolb2004}. These  close binaries are  commonly referred
to as  post-common envelope binaries  or PCEBs. Once they  are formed,
these objects are subject to  substantial angular momentum loss due to
gravitational  radiation and/or  magnetic  braking,  which make  their
orbital periods  shorter. Hence,  they are the  progenitors of  a wide
range  of  important  outcomes,   e.g.   type  Ia  supernovae,  double
degenerate  binaries,   white  dwarf  pulsars   \citep{Toloza2019}  or
highly-field  magnetic  white dwarfs  \citep{GarciaBerro2012}.   PCEBs
have played a  very important role in studying a  wide variety of open
problems  in   modern  astronomy.   These  include   constraining  the
efficiency  of  common  envelope   evolution  and  its  energy  budget
\citep{Zorotovic2010,   Rebassa2012,    Camacho2014,   Zorotovic2014},
confirming the  inefficiency of magnetic braking  for fully convective
stars \citep{Schreiber2010, Zorotovic2016}, proving that the origin of
low-mass ($\la0.45\,$M$_{\odot}$) white dwarfs is mainly a consequence
of  binary evolution  \citep{Rebassa2011}, constraining  the secondary
star  mass function  \citep{Ferrario2012,  Cojocaru2017},  as well  as
testing the mass-radius relation  of white dwarfs \citep{Parsons2017},
main sequence stars \citep{Parsons2018} and metal-poor sub-dwarf stars
\citep{Rebassa2019} via the analysis of eclipsing binaries.

\begin{figure*}
    \centering
    \includegraphics[width=\columnwidth]{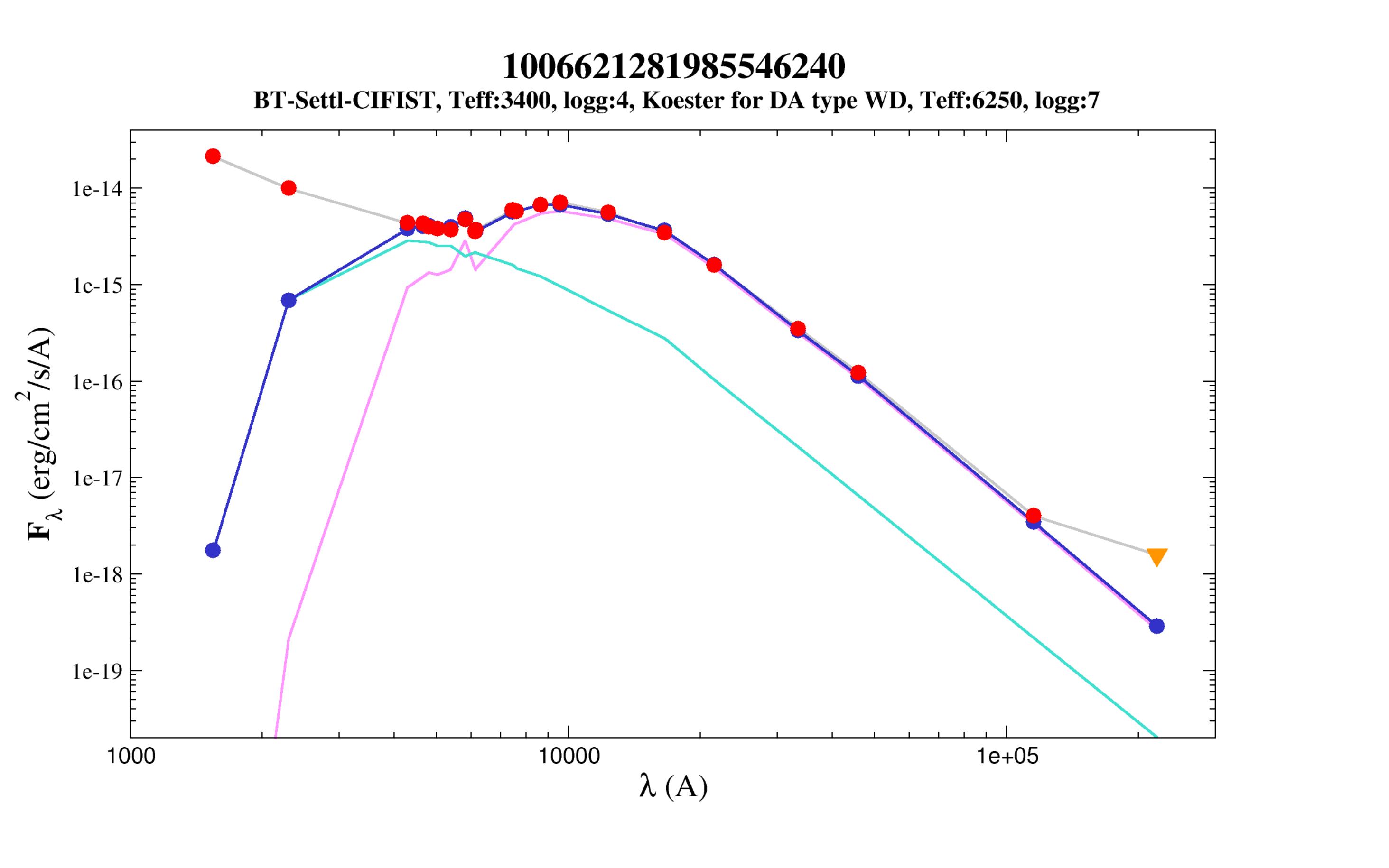}
    \includegraphics[width=\columnwidth]{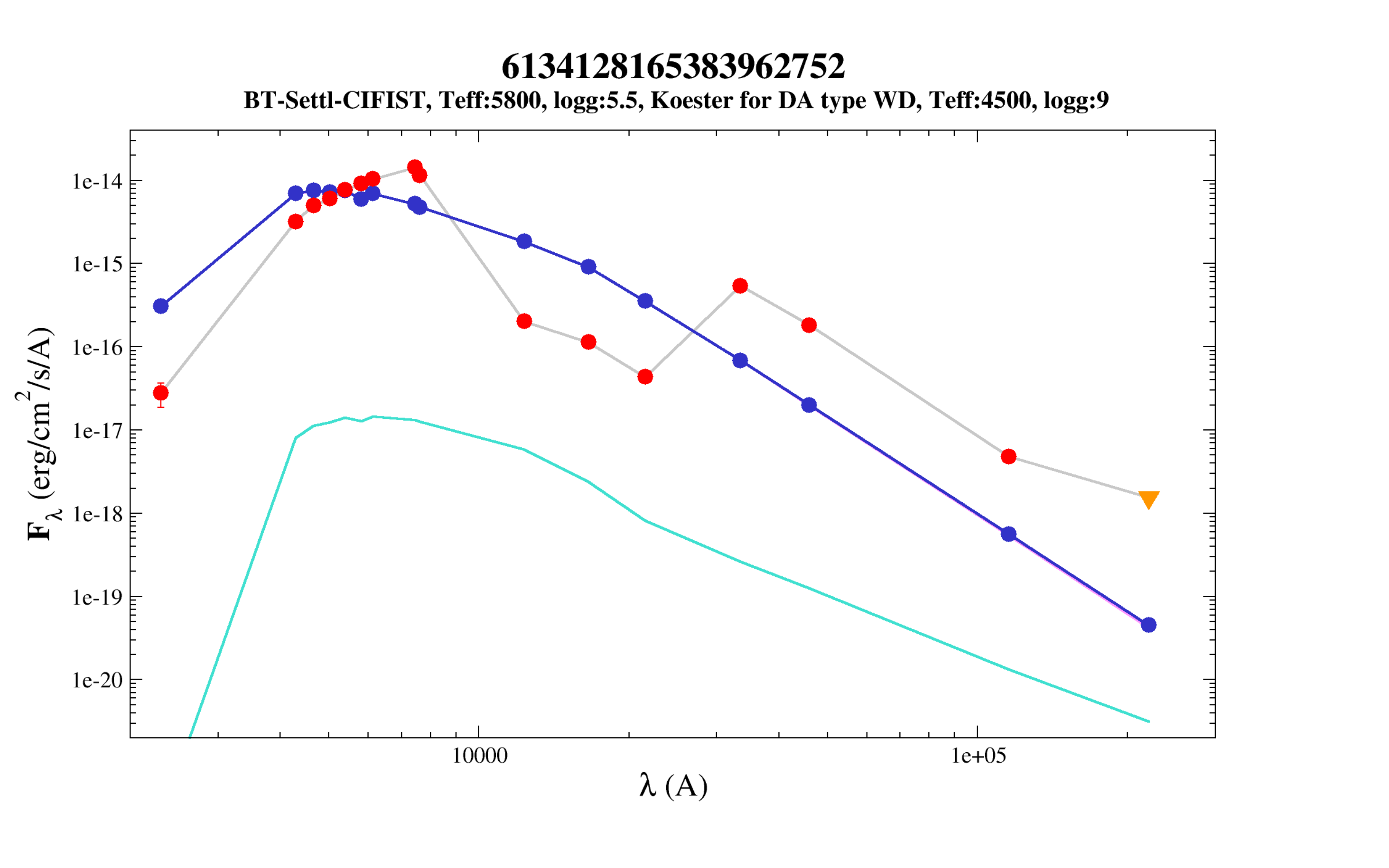}
    \includegraphics[width=\columnwidth]{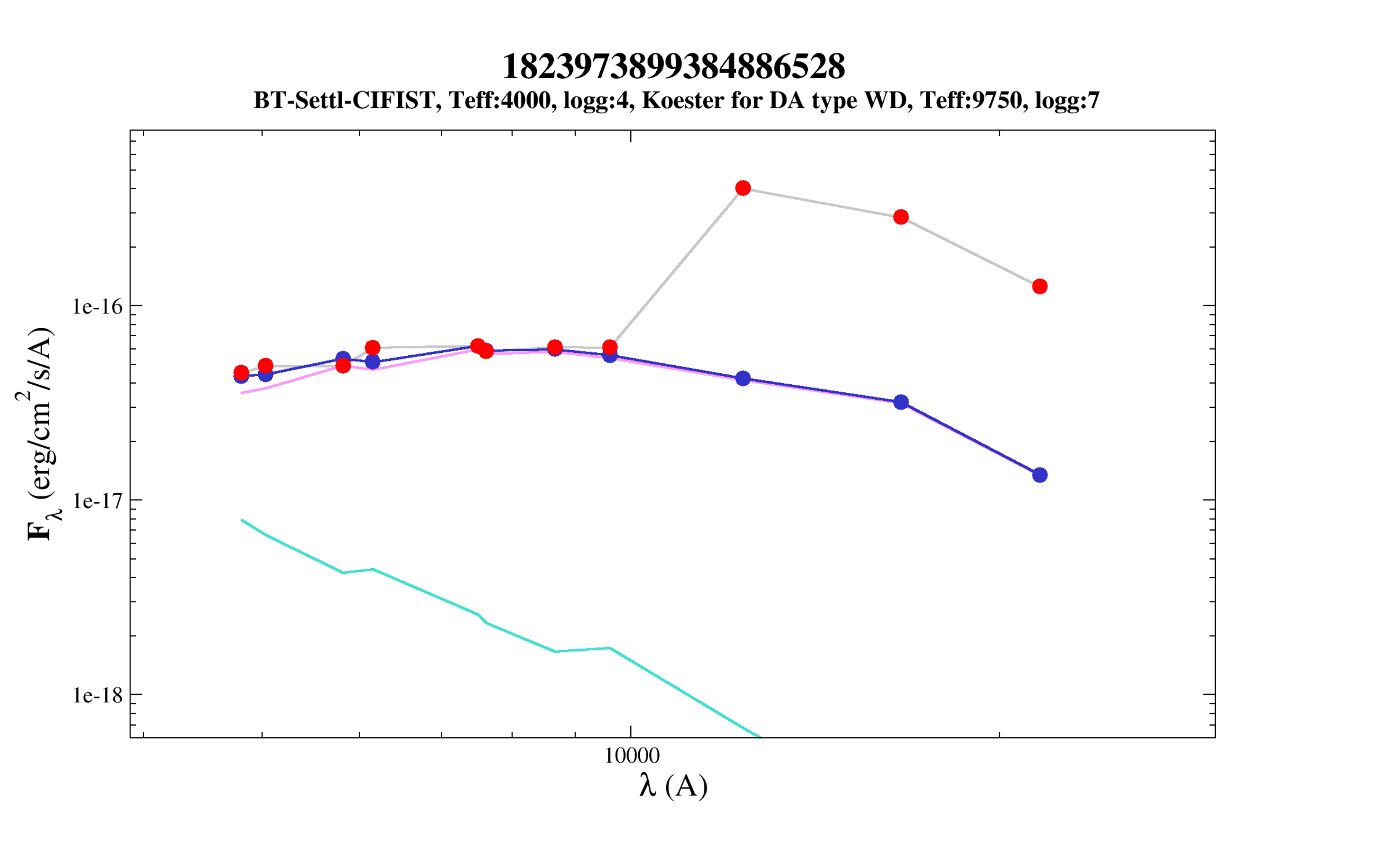}
    \caption{Different  examples of  bad  two-body  SED fittings  with
      VOSA. From left to right and from top to bottom: Bad UV fit, bad
      photometry,  and  contamination  from  a  nearby  star  at  WISE
      wavelengths.   The  cyan   and  magenta   lines  represent   the
      hydrogen-rich  white  dwarf  \citep{Koester10} and  CIFIST  main
      sequence star  \citep{Baraffe15} models, respectively.  Red dots
      are the observed photometric points while the blue line and dots
      indicate the composite model that best fits the data. The yellow
      inverted   triangles  indicate   that  the   photometric  values
      correspond to upper limits and, thus, are not taken into account
      in the fitting process.}
    \label{fig-VOSA_bad}
\end{figure*}

Behind  all the  important results  derived analysing  both wide  WDMS
binaries and PCEBs relies  a tremendous observational effort dedicated
to identifying  large and  homogeneous (i.e.  selected under  the same
criteria)  samples of  such  systems.  So far,  the  largest and  most
homogeneous  catalogues  of  spectroscopic  WDMS  binaries  have  been
obtained from  the Sloan Digital Sky  Survey (SDSS; \citealt{York2000,
  Eisenstein2011})  \citep{Rebassa2010,   Rebassa2012b,  Rebassa2013b,
  Rebassa2016b}   and   the   Large  Sky   Area   Multi-Object   Fiber
Spectroscopic Telescope  (LAMOST; \citealt{Cui2012,  Chen2012}) Survey
\citep{Ren2014, Ren2018}. Counting the SDSS sample with $\simeq$3\,200
systems  and   the  LAMOST  sample  with   $\simeq$900  objects,  both
catalogues have  been prolific at  revealing a large number  of radial
velocity    variable,   i.e.    PCEBs,   and    non-variable   systems
\citep{Rebassa2007,  Rebassa2011, Schreiber2008,  Schreiber2010} --and
the  corresponding  orbital period  distributions  \citep{Rebassa2008,
  Nebot2011} --  as well as of  eclipsing binaries \citep{Parsons2012,
  Parsons2013,  Parsons2015,   Pyrzas2009,  Pyrzas2012,  Rebassa2016b,
  Rebassa2019}.

However, it  is important to emphasise  that both the SDSS  and LAMOST
WDMS  binary  catalogues are  heavily  affected  by selection  effects
\citep{Rebassa2010, Ren2018}. For instance, being the SDSS a dedicated
survey  to  spectroscopically  follow-up  quasars  and  galaxies,  the
resulting WDMS  binary catalogue is biased  against the identification
of  cool white  dwarfs  (the dominant  underlying  population) and  is
dominated by  systems containing  hot white  dwarfs ($\ga$10\,000\,K),
which overlap  with quasars in  colour space. Cooler white  dwarfs are
also   under-represented  in   the   LAMOST   WDMS  binary   catalogue
\citep{Ren2018} because  of their intrinsic faintness.   Moreover, the
identification of both  components relies on the  visual inspection of
the  spectra, which  makes it  virtually impossible  to identify  WDMS
binaries  when  one  of  the   stars  dominates  the  spectral  energy
distribution (SED) at optical wavelengths.  As a consequence, the WDMS
binary  sample mainly  consists of  secondary stars  of spectral  type
M. Fortunately, a  step forward in this line has  been possible thanks
to the combination  of optical spectroscopy (from LAMOST  or RAVE, the
Radial  Velocity   Experiment;  \citealt{Kordopatis2013,  Kunder2017})
and/or  astrometry (from  TGAS, the  Tycho-Gaia astrometric  solution;
\citealt{Michalik2015} and \emph{Gaia}  DR2; \citealt{Evans2018}) with
ultraviolet  photometry   (from  GALEX,  Galaxy   Evolution  Explorer;
\citealt{Martin2005}), which has allowed  identifying large samples of
WDMS  binary  candidates  containing  earlier  type  A,  F,  G  and  K
companions  \citep{Parsons2016,  Rebassa2017, Ren2020,  Anguiano2020}.
Finally, the SDSS and  LAMOST catalogues are magnitude-limited samples
and, as  a consequence, it is  difficult to derive space  densities of
both wide WDMS binaries and PCEBs.

Thanks  to   the  parallaxes  provided  by   the  \emph{Gaia}  mission
\citep{Gaia2016,  Gaia2020} the  prospects of  obtaining statistically
large  and   homogeneous  volume-limited  samples   have  dramatically
increased.  For  example, the 100\,pc  volume-limited, nearly-complete
($\simeq$95  per  cent)  \emph{Gaia}  white dwarf  sample  consist  of
$\simeq$13\,000  objects \citep{Jimenez2018,  Gentile2019, Kilic2020},
which  is two  orders of  magnitudes larger  than previous  catalogues
\citep[e.g.][]{Giammichele2012, Holberg2016}. Comprehensive samples of
accreting  white  dwarfs  in cataclysmic  variables  \citep{Abril2020,
  Pala2020,   Abrahams2020},    extremely   low-mass    white   dwarfs
\citep{Pelisoli2019},  hot subdwarfs  \citep{Geier2019}, white  dwarfs
with   infrared  excess   typical   of   circumstellar  debris   disks
\citep{Rebassa2019a,  Xu2020}, and  white dwarfs  that are  members of
common proper motion pairs \citep{Elbadry2018, Elbadry2021}, have also
been  obtained from  analysing  the \emph{Gaia}  data. More  recently,
\citet{Inight2021}  have  illustrated  the importance  of  defining  a
volume-limited sample of close white dwarf binaries. Our motivation in
this work is  to identify a sample of unresolved  white dwarf binaries
within 100  pc, in particular  those having main  sequence companions,
i.e. WDMS. At this maximum distance  we expect the selected targets to
form  a nearly-complete  sample  \citep{Jimenez2018}. Moreover,  being
these WDMS binaries  unresolved in \emph{Gaia}, it is  expected that a
relatively large fraction of them should have evolved through a common
envelope phase.   This is because  the majority of wide  binaries that
evolved avoiding mass  transfer episodes are expected  to be partially
or fully  resolved at such  short distances.  Thus, by  analysing this
nearly-complete volume-limited sample  we are able to  derive an upper
estimate of the space density of PCEBs.  Unlike \citet{Belokurov2020},
who made use of the renormalised  unit weight error (RUWE) to identify
unresolved  binaries,  we  define specific  colour-magnitude  cuts  to
select unresolved  WDMS binary  candidates.  By applying  the two-body
fitting              algorithm             implemented              in
VOSA\footnote{\url{http://svo2.cab.inta-csic.es/theory/vosa/}}
(Virtual Observatory  SED Analyser; \citealt{Bayo08}) to  the selected
candidates, we are  able to exclude contaminants and  to determine the
white  dwarf and  main  sequence star  stellar  parameters, which  are
compared  to  those  obtained  from the  SDSS  WDMS  magnitude-limited
catalogue \citep{Rebassa2010, Rebassa2016b}.
 
\section{The selection of unresolved WDMS binaries}
\label{s-sample}

In the Hertzsprung-Russell (HR)  diagram, unresolved WDMS binaries are
expected to form a bridge between the main sequence star and the white
dwarf   star   loci   if   there  is   sufficient   flux   from   both
components. Otherwise, the flux of one  of the stars dominates the SED
and,  as  we have  already  mentioned,  these binaries  are  extremely
difficult to identify because they have similar magnitudes and colours
as single stars. We searched  for unresolved WDMS binary candidates in
which  the two  components  emit sufficient  flux  to be  individually
detected in the SEDs as follows.

First, we  derived synthetic WDMS  binary $G$ absolute  magnitudes and
$G_\mathrm{BP}-G_\mathrm{RP}$  colours.  We  did  this converting  the
\emph{Gaia} EDR3 absolute magnitudes that we incorporated\footnote{The
  synthetic   $G$,   $G_\mathrm{BP}$  and   $G_\mathrm{RP}$   absolute
  magnitudes were derived integrating the flux of the associated model
  atmosphere  spectra  \citep{Koester10}  rescaled at  a  distance  of
  10\,pc over the corresponding  EDR3 pass-bands and zero-points, that
  we                           obtained                           from
  \url{https://www.cosmos.esa.int/web/gaia/edr3-passbands}.}   in  the
hydrogen-rich  white  dwarf  cooling   sequences  of  La  Plata  group
\citep{Althaus2013,  Camisassa2016,   Camisassa2019}  to   fluxes  and
combining them  with the main  sequence star  (spectral types A  to M)
fluxes  that we  obtained  from transforming  the absolute  magnitudes
provided        by         the        updated         tables        of
\citet{Pecaut+Mamajek2013}\footnote{\url{http://www.pas.rochester.edu/~emamajek/EEM_dwarf\_UBVIJHK\_colors\_Teff.txt}}. We
then converted the  combined fluxes back into  absolute magnitudes and
thus obtained a grid of  synthetic WDMS binary $G$ absolute magnitudes
and $G_\mathrm{BP}-G_\mathrm{RP}$ colours.  The grid contained 28\,944
points with  white dwarf effective temperatures  and surface gravities
ranging  from   3\,000  to  100\,000\,K   and  from  7  to   9.5  dex,
respectively, and  main sequence  star effective  temperatures ranging
from 2\,400 to  9\,700\,K. As expected, those  synthetic WDMS binaries
with  a  white  dwarf  (main sequence  star)  dominant  component  had
associated magnitudes nearly identical to those of single white dwarfs
(main sequence stars).

Second, we implement  a set of cuts for selecting  WDMS binaries based
on  the  position  of  the   synthetic  $G$  absolute  magnitudes  and
$G_\mathrm{BP}-G_\mathrm{RP}$   colours  in   the   HR  diagram   (see
Fig.\,\ref{fig-HR}, left panel).  These cuts were developed to exclude
single white dwarfs and main sequence stars (unfortunately, these also
excluded WDMS binaries  in which one of the stars  dominates the SED),
thus selecting WDMS binary candidates  in which the two components are
expected to contribute  to the SEDs.  We note that  the synthetic main
sequence  absolute magnitudes  are  provided for  the \emph{Gaia}  DR2
filters by \citet{Pecaut+Mamajek2013}. However, as  can be seen in the
left  panel of  Figure\,\ref{fig-HR}, the  position of  synthetic WDMS
binaries with  negligible white dwarf fluxes  perfectly coincides with
the main loci of observed EDR3  single main sequence stars.  Since the
main purpose  here is  to apply  cuts for  excluding single  stars, we
consider  that using  synthetic  \emph{Gaia} DR2  magnitudes for  main
sequence stars has no major impact in our analysis.

\begin{figure}
    \centering
    \includegraphics[width=\columnwidth]{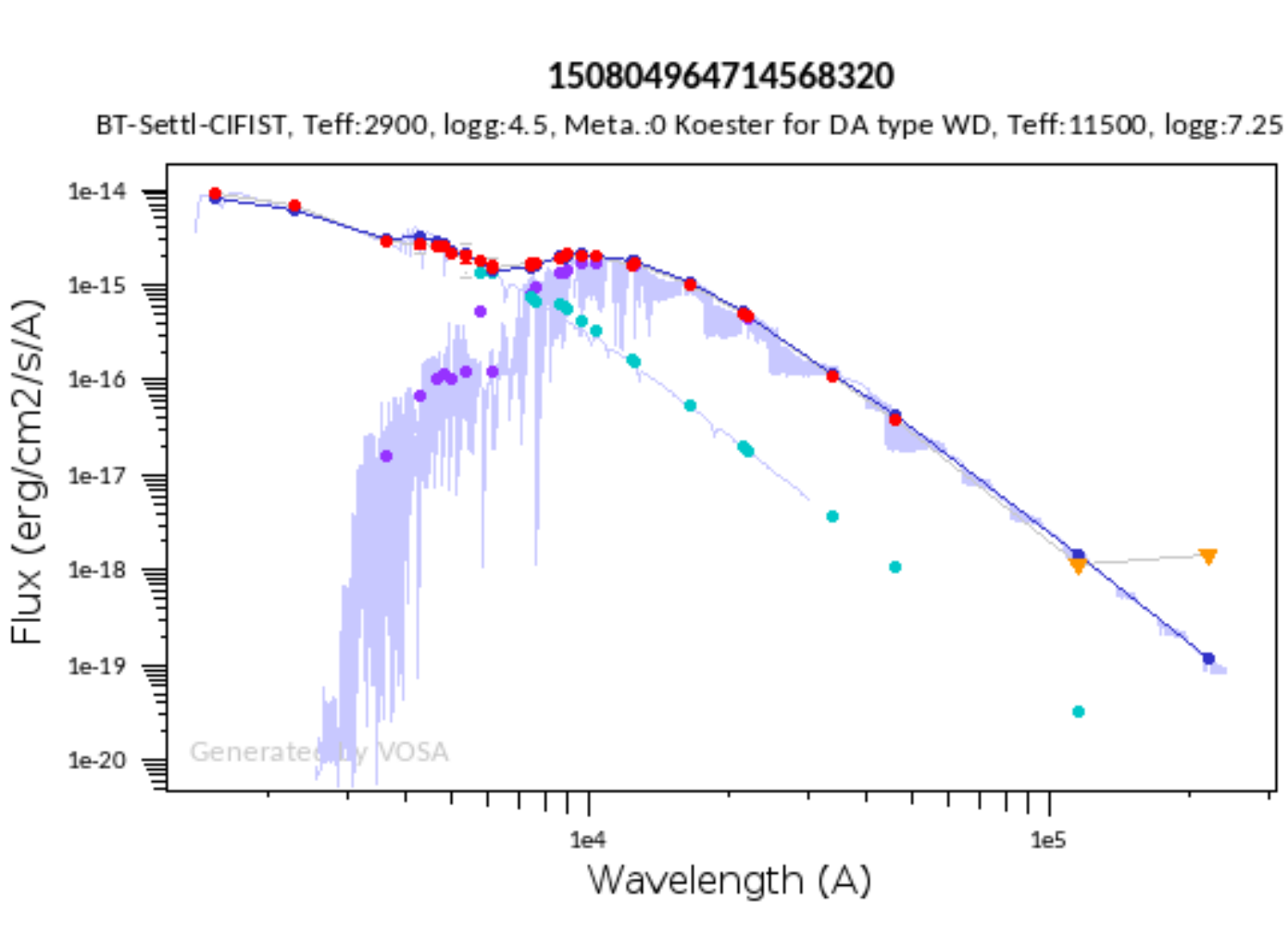}
    \caption{Composite    SED    of   the    source    Gaia\_EDR3\_ID:
      150804964714568320. The  cyan dots and the  corresponding purple
      line represent the hydrogen-rich white dwarf Koester model while
      the purple dots  and the corresponding purple  line indicate the
      CIFIST  model.   The  rest   of  symbols   and  colours   as  in
      Fig.\,\ref{fig-VOSA_bad}.}
    \label{fig-VOSA}
\end{figure}

The  cuts for  selecting  WDMS binary  candidates  are illustrated  in
Figure\,\ref{fig-HR} as red solid lines and follow these equations:

\begin{eqnarray}
[(M_\mathrm{G} > 5.76332-1.00990(G_\mathrm{BP}-G_\mathrm{RP})\nonumber \\-0.78017(G_\mathrm{BP}-G_\mathrm{RP})^2+4.21489(G_\mathrm{BP}-G_\mathrm{RP})^3\nonumber\\+1.71673(G_\mathrm{BP}-G_\mathrm{RP})^4-25.49710(G_\mathrm{BP}-G_\mathrm{RP})^5\nonumber\\-13.68387(G_\mathrm{BP}-G_\mathrm{RP})^6+24.16992(G_\mathrm{BP}-G_\mathrm{RP})^7\nonumber\\ \hspace{0.2cm} \mathrm{AND} \hspace{0.2cm} -0.65 < (G_\mathrm{BP}-G_\mathrm{RP}) < 0.58)\nonumber\\
\mathrm{OR} \nonumber \\
(M_\mathrm{G} > 21.22759-118.35415(G_\mathrm{BP}-G_\mathrm{RP})\nonumber\\+305.43033(G_\mathrm{BP}-G_\mathrm{RP})^2-393.13909(G_\mathrm{BP}-G_\mathrm{RP})^3\nonumber\\+294.22156(G_\mathrm{BP}-G_\mathrm{RP})^4-135.40159(G_\mathrm{BP}-G_\mathrm{RP})^5\nonumber\\+38.90143(G_\mathrm{BP}-G_\mathrm{RP})^6-6.80771(G_\mathrm{BP}-G_\mathrm{RP})^7\nonumber\\+0.66434(G_\mathrm{BP}-G_\mathrm{RP})^8-0.02773(G_\mathrm{BP}-G_\mathrm{RP})^9 \nonumber \\\hspace{0.2cm} \mathrm{AND} \hspace{0.2cm} 0.58 \leq (G_\mathrm{BP}-G_\mathrm{RP}) < 4.5)]
\end{eqnarray}

AND

\begin{eqnarray}
[(M_\mathrm{G} < 11.13367+3.51045(G_\mathrm{BP}-G_\mathrm{RP})\nonumber\\-1.13132(G_\mathrm{BP}-G_\mathrm{RP})^2+0.30628(G_\mathrm{BP}-G_\mathrm{RP})^3\nonumber\\ \hspace{0.2cm} \mathrm{AND} \hspace{0.2cm} -0.65 < (G_\mathrm{BP}-G_\mathrm{RP}) < 1.7)\nonumber\\
\hspace{0.2cm} \mathrm{OR} \hspace{0.2cm}\nonumber\\
(M_\mathrm{G} < 14.587+0.45098(G_\mathrm{BP}-G_\mathrm{RP})\nonumber\\ \hspace{0.2cm} \mathrm{AND} \hspace{0.2cm} 1.7 \leq (G_\mathrm{BP}-G_\mathrm{RP}) < 4.25)\nonumber\\
\hspace{0.2cm} \mathrm{OR} \hspace{0.2cm}\nonumber\\
(M_\mathrm{G} < 40.3-5.6(G_\mathrm{BP}-G_\mathrm{RP})\nonumber\\ \hspace{0.2cm} \mathrm{AND} \hspace{0.2cm}  4.25 \leq (G_\mathrm{BP}-G_\mathrm{RP}) < 4.5))]
\end{eqnarray}

We  used  the above  cuts  to  select  WDMS binary  candidates  within
\emph{Gaia}  EDR3,  together with  the  following  conditions that  we
applied  in  order to  ensure  good  quality  in the  photometric  and
astrometric values:

\begin{itemize}
\item 1/$\varpi \leq$ 100 
\item $\varpi/\sigma_{\varpi} \ge$ 10 
\item $I_{\rm BP}/\sigma_{I_{\rm BP}} \ge$ 10
\item $I_{\rm RP}/\sigma_{I_{\rm RP}} \ge$ 10
\item $I_{\rm G}/\sigma_{I_{\rm G}} \ge$ 10
\end{itemize}

\noindent where $\varpi$  is the parallax in  arcseconds, $I_{\rm G}$,
$I_{\rm BP}$ and  $I_{\rm RP}$ are the fluxes in  the bandpass filters
$G$, $G_\mathrm{BP}$ and $G_\mathrm{RP}$ respectively and the $\sigma$
values are the  standard errors of the  corresponding parameters. This
resulted in 20\,719  WDMS binary candidates, which  are illustrated in
the  right panel  of  Figure\,\ref{fig-HR}.  As it  can  be seen  from
inspection of this  Figure, there seems to be a  large number of false
positive candidates, mainly due to the  fact that we did not apply any
condition on the excess flux factor. While the flux in the $G$ band is
determined  from  a profile-fitting,  the  $BP$  and $RP$  fluxes  are
computed  by  summing  the  flux  in  a  field  of  3.5  $\times$  2.1
arcsec$^{2}$    \citep{Evans2018}.     Given    the    $G_\mathrm{BP},
G_\mathrm{RP}$  and  $G$ passbands,  the  sum  of $G_\mathrm{BP}$  and
$G_\mathrm{RP}$  fluxes should  exceed the  $G$ flux  by only  a small
factor.   Larger deviations  should  be caused  by contamination  from
nearby  sources  or  bright  background  in  the  $G_\mathrm{BP}$  and
$G_\mathrm{RP}$ bands.

In order to quantify this effect, \citet{Evans2018} defined the excess
factor as the  flux ratio\, C = ($I_\mathrm{BP}$  +$ I_\mathrm{RP}$) /
$I_\mathrm{G}$ . However, as \citet{Riello20}  pointed out, there is a
clear colour  dependence of the excess  factor that is not  taken into
account in the  previous expression. To overcome  this situation, they
introduced  a  corrected  excess  factor  defined as  C$^{*}$  =  C  -
f($G_\mathrm{BP},     G_\mathrm{RP}$)      where     f($G_\mathrm{BP},
G_\mathrm{RP}$) is a function providing the expected excess at a given
colour.   According to  this,  a  C$^{*}$ value  close  to zero  would
indicate  that the  source  is  not affected  by  excess  flux in  the
$G_\mathrm{BP}$ or $G_\mathrm{RP}$ band.

Using C$^{*}$ and  adequate values of sigma -- the  scatter of C$^{*}$
as a function of the $G$  magnitude, see Eq. 18 in \citet{Riello20} --
we ended up with 2\,001  sources with consistent photometric values in
the  $G_\mathrm{BP}, G_\mathrm{RP}$  and $G$  filters. It  is possible
that a fraction of the objects  we have excluded because of the excess
factor    criteria   are    real   but    partially   resolved    WDMS
binaries.  However, we  emphasise that  the goal  of this  work is  to
identify unresolved WDMS binaries.

\section{Stellar parameters of the final unresolved WDMS binary sample}
\label{s-param}

We  executed VOSA  to estimate  the stellar  parameters of  the 2\,001
candidates selected in the previous Section. Firstly, we built the SED
of the objects  from the UV to the mid-IR  using the public catalogues
accessible   within  VOSA.    In  particular   we  used   GALEX  GR6+7
\citep{Bianchi17},    APASS    DR9   \citep{Henden15},    SDSS    DR12
\citep{Alam15},  Pan-Starrs  PS1   DR2  \citep{Magnier20},  Gaia  EDR3
\citep{Gaia21,  Gaia16}, 2MASS  PSC  \citep{Skrutskie06}, and  AllWISE
\citep{Wright10}.  VOSA  gathers not only the  photometric information
(magnitudes and  errors) but also  the associated quality  flags. This
way, photometric points with poor  quality were removed before the SED
fitting.  Also, upper limits were not considered either.

Given  that  our targets  are  close  objects ($\leq100$\,pc),  proper
motions  are expected  to  be  high.  We  took  them  into account  by
computing the coordinates  of the targets in the epoch  2000 using the
astrometric information provided by \emph{Gaia},  and then applied a 5
arcsec search radius around the epoch 2000 coordinates.

For 70 sources,  the number of collected  photometric measurements was
too small  to carry out the  SED fitting.  After visual  inspection we
discarded another  302 sources  having wrong photometric  values.  The
rest (1\,629) were fitted using either the main sequence star BT-Settl
CIFIST   \citep{Baraffe15}   or    the   hydrogen-rich   white   dwarf
\citep{Koester10} collections of theoretical models.  In total, 1\,349
objects  had a  good SED  fitting (\emph{vgfb}$<$15),  indicating that
their  SEDs  resemble  those  of single  stars\footnote{This  list  of
  excluded   objects  presumably   includes  also   unresolved  double
  degenerates  that  passed  our  selection cut  and  which  had  SEDs
  virtually identical  to those of single  white dwarfs.}. \emph{vgfb}
is a modified $\chi^2$, internally used by VOSA, that is calculated by
forcing  $\Delta   F_\mathrm{i}$  to   be  larger  than   0.1  $\times
F_\mathrm{i}$,  where  $\Delta  F_\mathrm{i}$  is  the  error  in  the
observed  $F_\mathrm{i}$  for  the  $i$-th  flux  in  the  SED.   This
parameter is particularly useful when the photometric errors of any of
the   catalogues   used  to   build   the   SED  are   underestimated.
\emph{vgfb}$<$ 15 is a reliable  indicator of good fit.  The remaining
280 (1\,629-1\,349) are candidates to  WDMS binary systems.  For these
objects  we  took  advantage  of  the  two-body  fitting  capabilities
implemented in VOSA. 129 objects could not be fitted due to, at least,
one of  the following three reasons:  UV excess likely due  to stellar
activity  of  the   main  sequence  companion,  bad   SED  fitting  or
contamination from a  nearby star, in particular  at WISE wavelengths.
In  Figure\,\ref{fig-VOSA_bad} we  show  an example  for  each of  the
previous cases. In order to ensure  that our sample is actually within
100  pc from  the Sun,  we removed  from the  remaining 151  (280-129)
objects     those     having     bad     astrometry     (RUWE>2     or
astrometric$_{-}$excess$_{-}$noise>2                               and
astrometric$_{-}$excess$_{-}$noise$_{-}$sig>2).  After  this, we ended
up with  a final list of  117 good candidates to  WDMS binaries within
100\,pc. An  example of  the SED  fitting of one  of these  sources is
shown in Figure\,\ref{fig-VOSA}.

\begin{figure}
    \centering
    \includegraphics[angle=-90,width=\columnwidth]{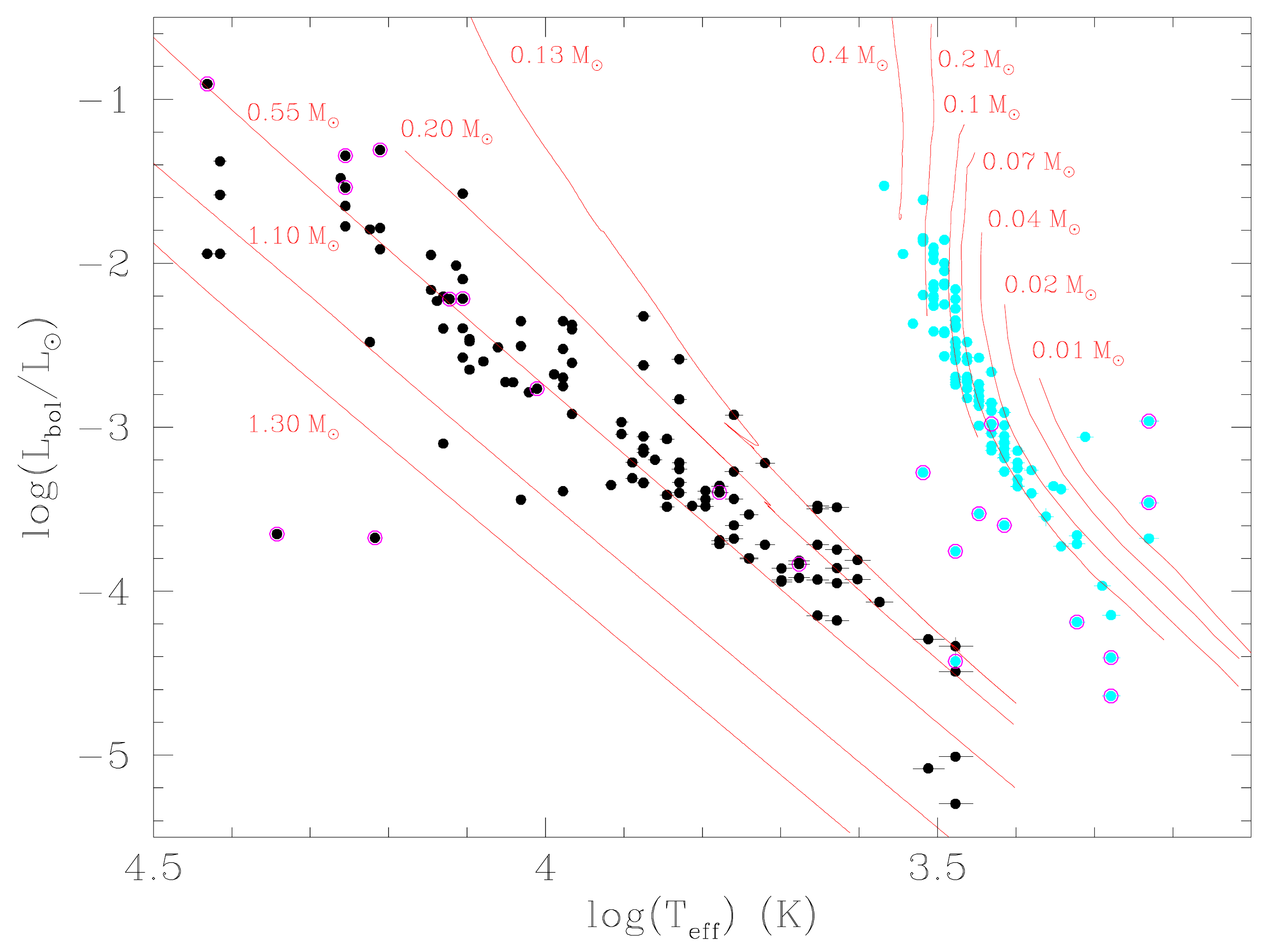}
    \caption{Luminosities as a function  of effective temperatures for
      the 117  white dwarfs  (black solid  dots) and  their companions
      (cyan  solid  dots) that  are  part  of unresolved  WDMS  binary
      candidates and with good two-body VOSA fits. The solid red lines
      show  the hydrogen-rich  white dwarf  theoretical relations  for
      fixed   masses   of  0.13\,M$_{\odot}$   and   0.20\,M$_{\odot}$
      \citep{Calcaferro2018},  0.55\,M$_{\odot}$ \citep{Camisassa2016}
      and 1.1\,M$_{\odot}$  and 1.3\,M$_{\odot}$\citep{Camisassa2019},
      the  main sequence  theoretical relations  for a  fixed mass  of
      0.40\,M$_{\odot}$,   0.20\,M$_{\odot}$,  0.10\,M$_{\odot}$   and
      0.07\,M$_{\odot}$   \citep{Baraffe15}   and  the   brown   dwarf
      theoretical  relations for  0.04\,M$_{\odot}$, 0.02\,M$_{\odot}$
      and 0.01\,M$_{\odot}$  (CIFIST models).  The pairs  for which at
      least  one  of  the   components  has  a  temperature-luminosity
      observed  relation  which  is  far  away  from  the  theoretical
      expected values are surrounded by magenta open circles.}
    \label{fig-lbol-teff}
\end{figure} 

For the 117 unresolved WDMS  binary candidates with good two-body fits
VOSA   provides  information,   for  each   component,  on   effective
temperatures   ($T_{\rm    eff}$)   and   total    bolometric   fluxes
($F_\mathrm{bol}$). The temperatures are estimated from the model that
best fits  the data after the  appropriate re-scaling in flux.   For a
detailed description on how the  total bolometric flux is computed see
the                                                               VOSA
documentation\footnote{\url{http://svo2.cab.inta-csic.es/theory/vosa/helpw4.php?otype=star&what=intro}}. Luminosities
are then obtained from the equation
\begin{equation}
    L_\mathrm{bol}(L_{\odot}) = 4\pi D^2 F_\mathrm{bol}
\end{equation}
where  $L_\mathrm{bol}$  is  the  bolometric  luminosity  (luminosity,
hereafter),  that  is,  the  total amount  of  electromagnetic  energy
emitted per  unit of time and  $D$ represents the distance  in parsecs
obtained from  the inverse of the  \emph{Gaia} parallax\footnote{Given
  that   our    sources   are    not   farther   than    100\,pc   and
  $\varpi/\sigma_{\varpi}   \ge$  10,   no  further   corrections  are
  needed.}.  In Figure\,\ref{fig-lbol-teff} we show these luminosities
as  a  function  of  the   effective  temperatures  derived  from  the
best-model fits for  both the white dwarfs and  their companions.  For
comparison, we also represent  the theoretical $L_\mathrm{bol}$ versus
$T_\mathrm{eff}$ relations for different white dwarf and main sequence
star and brown  dwarf masses. As it  can be seen from  the figure, the
$L_\mathrm{bol}$ and $T_\mathrm{eff}$ values  measured by VOSA fall in
the expected region for the white dwarfs, except in two cases in which
the luminosities are too low  for the expected effective temperatures.
For a companion to one of these two white dwarfs, as well as in 9 more
cases, the  observed parameters  of the main  sequence stars  are also
clearly   away    from   the   expected   theoretical    values   (See
Figure\,\ref{fig-lbol-teff}).   One  possible   explanation  for  this
discrepancy is  the existence  of another  component, for  instance an
accretion disk as in cataclysmic  variables, which has not been taking
into  account  in  our  analysis.  Consequently,  given  the  lack  of
spectroscopy  we cannot  confirm or  disprove the  nature of  these 11
objects to be real WDMS binaries  and therefore we decided to consider
them    as    candidates    with   unreliably    determined    stellar
parameters. These objects  are represented by open  magenta circles in
Figure\,\ref{fig-lbol-teff}.

Radii  are  estimated by  VOSA  using  the following  two  independent
equations for each component:

\begin{equation}
   R_1 = \sqrt{D^2 M_\mathrm{D}} \\
\end{equation}
where M$_\mathrm{D}$ is the scaling factor  of the model fluxes to fit
the observed ones, and \\
\begin{equation}
    R_2 = \sqrt{L_\mathrm{bol}/(4\pi\sigma_{SB} T_\mathrm{eff}^4)}
\end{equation}
where $\sigma_{SB}$ is the Stefan-Boltzmann  constant, with a value of
5.67~$\times$10$^{-8}$~W\,m$^{-2}$\,K$^{-4}$. Both  expressions return
almost identical  values for  the radii and  we adopt  those resulting
from Eq. (5).

\begin{figure*}
    \centering
    \includegraphics[width=0.8\textwidth]{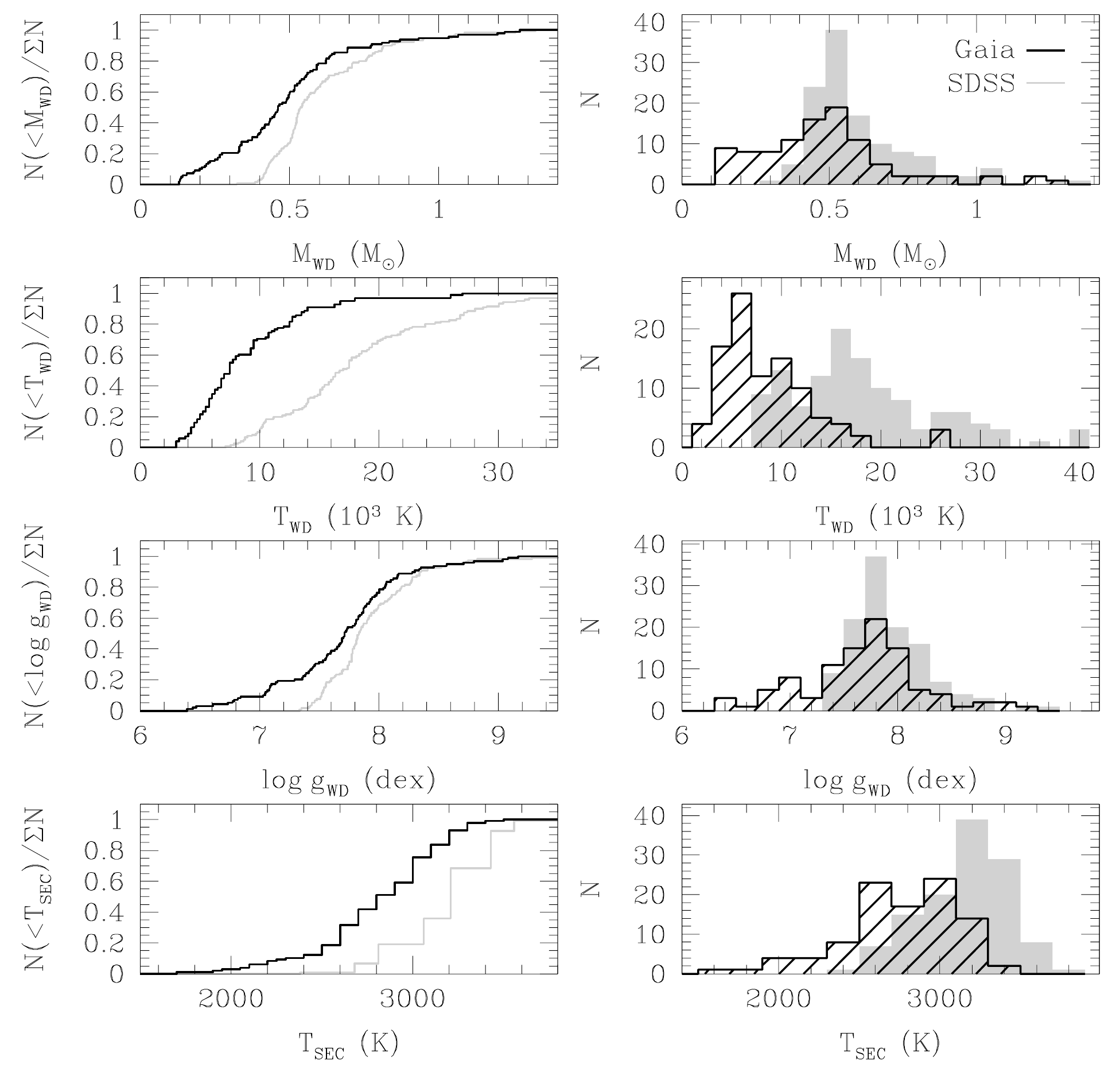}
    \caption{Comparison  between the  white  dwarf stellar  parameters
      (mass, effective temperature and  surface gravity) and secondary
      star effective  temperatures derived for the  WDMS components in
      the \emph{Gaia} (black) and SDSS (gray) samples. The left panels
      show the  cumulative distributions; the right  panels the normal
      distributions.}
    \label{fig-comp}
\end{figure*} 

Once the effective  temperatures and radii of  both stellar components
are  determined from  the VOSA  two-body  SED fitting,  we obtain  the
surface   gravities  of   the  white   dwarfs  by   interpolating  the
corresponding $T_\mathrm{eff}$  and $R$ values in  the (hydrogen-rich)
cooling  sequences  of the  La  Plata  group: \citet{Althaus2013}  and
\citet[][we    adopted   the    tracks    with    a   thin    hydrogen
  envelope]{Calcaferro2018}  for   low-mass  and   extremely  low-mass
He-core white dwarfs,  respectively, \citet{Camisassa2016} for CO-core
white dwarfs  and \citet{Camisassa2019} for ONe-core  white dwarfs. It
is  worth  mentioning that  the  evolutionary  sequences of  extremely
low-mass white dwarfs are rather complicated, as these objects undergo
hydrogen flash phases that move them back and forth and up and down in
the  $\log(L_\mathrm{bol})$-$\log(T_\mathrm{eff})$  diagram  \cite[see
  e.g.  Figure 1 of][]{Istrate2016}.  We assume our extremely low-mass
white dwarfs  to have already  passed these flashes  and to be  in the
final stage of their cooling, which is a good assumption given the low
effective temperatures and luminosities  measured by the VOSA fitting.
The  white dwarf  masses  are  then easily  derived  from the  surface
gravity and  radius determinations.  For three  WDMS binary candidates
(IDs 2630815357409558400, 4550719958390541824 and 2027615341341444608)
the  white  dwarf   masses  were  estimated  to  be   lower  than  the
0.13$\,M_{\odot}$ lowest limit from our theoretical models.

We cross-matched our  117 \emph{Gaia} WDMS binary  candidates with the
SIMBAD  database \citep{Wenger00}  in  order to  identify the  already
known sources and collect information  about their nature. For this we
computed the coordinates  of the targets in the epoch  J2000 using the
astrometric information provided by \emph{Gaia},  and then applied a 3
arcsec search  radius around the  epoch J2000 coordinates. Of  the 117
WDMS candidates, 20  were already identify as binary  systems, five of
them  as cataclysmic  variables  (Table\,\ref{tab:CV};  note that  for
these targets  the luminosities and effective  temperatures derived by
VOSA, also included in the table,  were in agreement with the expected
theoretical  values  in  Figure\,\ref{fig-lbol-teff}).  Two  of  these
cataclysmic    variables    (Gaia   EDR3    6185040879503491584    and
783921244796958208)  were  also  included  in  \citet{Pala2020}.   The
remaining  97 objects  are hence  new identifications  as WDMS  binary
candidates.

\begin{table*}
\caption{\label{tab:CV}  List  of  cataclysmic variables  reported  in
  SIMBAD and included in our list  of 117 WDMS binary candidates.  The
  last   two  columns   indicate   the   luminosities  and   effective
  temperatures derived by  VOSA for the white dwarf  and main sequence
  star components.}
\setlength{\tabcolsep}{2ex}
\begin{center}
\begin{tabular}{ccccccc} 
\hline
\emph{Gaia} ID EDR3 & RA & DEC & log($L_\mathrm{bol}$)$_\mathrm{WD}$ & log($T_\mathrm{eff}$)$_\mathrm{WD}$ & log($L_\mathrm{bol}$)$_\mathrm{MS}$ & log($T_\mathrm{eff}$)$_\mathrm{MS}$ \\
          & (J2016)  & (J2016)  & ($L_{\odot}$)                 &      (K)              &   ($L_{\odot}$)   & (K)  \\
\hline
5034416735723964800 & 14.87104 & -26.51935  &  3.86 & -3.20 & 3.48 & -2.693 \\ 		
6185040879503491584 & 193.10032 & -29.24875 &  4.11 & -1.57 & 3.50 & -2.416 \\
674214551557961984 & 118.77167 & 22.00122   &  4.26 & -1.48 & 3.57 & -1.527 \\
755705822218381184 & 156.61466 & 38.75055   &  4.13 & -3.10 & 3.50 & -2.259 \\
783921244796958208 & 168.93529 & 42.97289   &  4.41 & -1.58 & 3.48 & -2.547 \\
\hline
\end{tabular}
\end{center}
\end{table*}

For the  15 (20-5) objects  reported in  SIMBAD as WDMS,  the spectral
types of  the M dwarf companions  are provided for three  of them. The
temperatures obtained by VOSA for these sources were in agreement with
the  tabulated  spectral types  by  SIMBAD,  using the  correspondence
between spectral  type and  effective temperature  for dwarf  stars in
\cite{Pecaut+Mamajek2013}.

We also cross-matched  our \emph{Gaia} sample of  WDMS candidates with
the SDSS WDMS binary catalogue within 100 pc. Seven SDSS WDMS binaries
fulfilled all the criteria used  to select our \emph{Gaia} sample. All
of  them were  identified  by our  methodology. In  the  same way,  we
checked if the 13  PCEBs from \citet{Schreiber+gaensicke2003} that are
presumably located below  100\,pc were in our list.  Of  these, 6 were
not   included  in   our   original  selection   of  20\,179   objects
(Section\,\ref{s-sample});  4 because  the \emph{Gaia}  EDR3 distances
are actually larger than 100\,pc and  the other 2 because they are not
within our WDMS region in the HR diagram (they are located on the main
sequence). The  remaining 7 objects  were picked up by  our selection,
however one of  them had bad photometry and three  more had bad 2-body
VOSA      fits     and      were      therefore     excluded      (see
Figure\,\ref{fig-VOSA_bad}).

Finally, we searched  for available public spectra of  our 117 sources
and  found   28  matches.    The  spectra   were  obtained   from  the
SDSS\footnote{\url{http://skyserver.sdss.org/dr16/en/tools/search/SQS.aspx}}
archive                  and                 from                  the
LAMOST\footnote{\url{ivo://org.gavo.dc/lamost6/q/svc$_{-}$lrs}}    and
ESO\footnote{\url{ivo://eso.org/ssap}}  VO  services   using  the  SVO
Discovery
Tool\footnote{\url{sdc.cab.inta-csic.es/SVODiscoveryTool}}. Inspection
of the spectra  and their SEDs revealed all our  targets to be genuine
WDMS binaries  except one  system which  turned out to  be one  of the
known cataclysmic variables included in Table\,\ref{tab:CV} (Gaia EDR3
783921244796958208).

After excluding the identified cataclysmic variables, our final sample
of unresolved  \emph{Gaia} WDMS  binary candidates  is reduced  to 112
(117-5) sources  and for  98 of  them (excluding  the 11  magenta open
circle  pairs  in Fig.\,\ref{fig-lbol-teff}  and  the  3 systems  with
estimated white dwarf masses lower than 0.13\,$M_{\odot}$) the stellar
parameters are considered to be  reliable. The stellar parameters, the
SED  two-body fits  and the  information for  each of  these 112  WDMS
binary candidates can be found at the SVO archive of WDs (see Appendix
A).

\section{Comparison between the SDSS and the \emph{Gaia} WDMS binary samples}

In the previous section we  have identified 112 unresolved \emph{Gaia}
WDMS  binary  candidates within  100  pc  and measured  their  stellar
parameters.  For  98 of  them we  found the  derived parameters  to be
consistent  with the  theoretical  expectations. Here  we compare  the
resulting stellar parameter distributions  of these 98 objects (namely
the white dwarf effective  temperatures, surface gravities and masses,
and  secondary   star  effective  temperatures)  to   those  from  the
magnitude-limited  SDSS  sample. We  note  that  in this  exercise  we
considered  SDSS WDMS  binaries  consisting of  a hydrogen-rich  white
dwarf and  a M dwarf companion,  since these are the  only systems for
which   the   full   set    of   stellar   parameters   is   available
\citep{Rebassa2010}.   Moreover,   we  excluded  SDSS   WDMS  binaries
displaying no  significant radial velocity variations,  i.e non-PCEBs.
This was  required since the  majority of unresolved  \emph{Gaia} WDMS
binaries  have evolved  through a  common envelope  phase and  are now
close     compact     binaries     (see     further     details     in
Section\,\ref{s-compl}), and  the properties  of close PCEBs  and wide
WDMS are statistically different \citep{Rebassa2011}.

The right panels of  Figure\,\ref{fig-comp} show the stellar parameter
distributions arising  from the two samples.  It has to be  noted that
spectral types, and not effective temperatures, are provided for the M
dwarf  companions of  the SDSS  sample.  For a  proper comparison,  we
converted these  into effective temperatures using  the updated tables
of \citet{Pecaut+Mamajek2013}.  Inspection of the figure  reveals that
the  properties  of  \emph{Gaia}  WDMS  binaries  seem  to  be  rather
different than those  of the SDSS ones, as expected  from a comparison
between  a   volume-limited  and   a  magnitude-limited   sample.  The
cumulative  stellar  parameter  distributions  for  both  samples  are
illustrated  in  the left  panels  of  Figure\,\ref{fig-comp}. We  run
Kolmogorov-Smirnov (KS) tests to  the white dwarf cumulative parameter
distributions to  evaluate the  probability that  the two  samples are
drawn  from  the same  parent  population.  Since the  secondary  star
effective temperature values  are discrete for both samples,  we run a
$\chi^2$-test  in this  case. Not  surprisingly, the  probabilities we
obtained  were  of the  order  of  less than  $10^{-4}$  ($>3.5\sigma$
significance) in all cases.

To begin with,  the \emph{Gaia} sample contains  a considerably larger
fraction of cool ($<10\,000$ K) white dwarfs. Such white dwarfs, being
fainter,  are  intrinsically  less  numerous  in  a  magnitude-limited
sample. Moreover, as we have already  mentioned, most of the SDSS WDMS
binaries were  observed because of  their similar colours  to quasars,
which overlap  in colour  space especially when  the white  dwarfs are
hotter than 10\,000 K. The fact that \emph{Gaia} WDMS binaries contain
systematically  cooler  white  dwarfs   was  already  pointed  out  by
\citet{Inight2021}.

\begin{figure}
    \centering
    \includegraphics[width=\columnwidth]{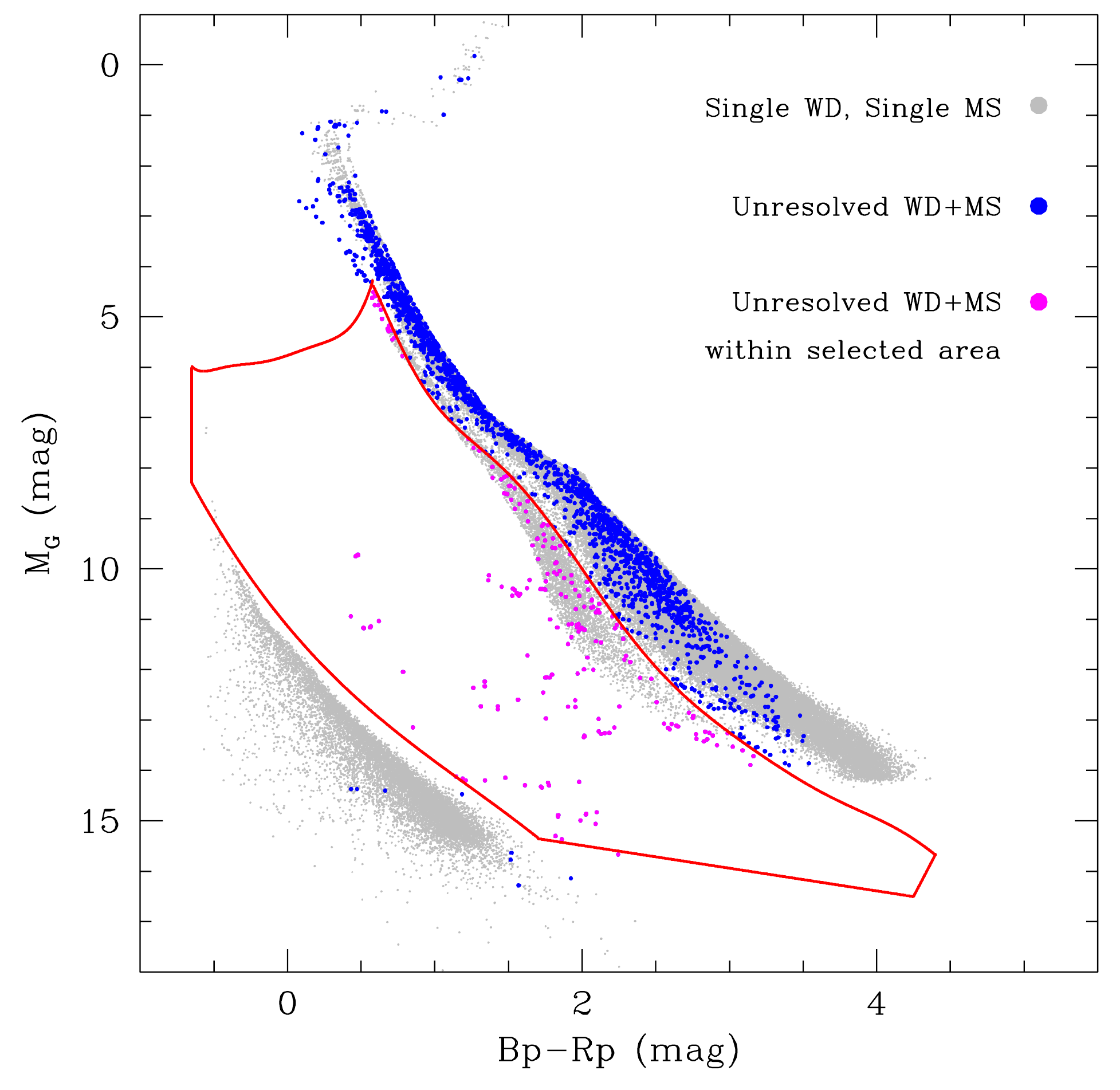}
    \caption{Simulated HR  diagram displaying single white  dwarfs and
      single main  sequence stars  (grey solid dots),  unresolved WDMS
      binaries (blue solid dots)  and unresolved WDMS binaries passing
      our selection cuts (solid magenta  dots). The selection cuts are
      illustrated as red solid lines.}
    \label{fig-HR-simul}
\end{figure} 

The white dwarfs  in both samples (volume-  and magnitude-limited) are
systematically less massive than canonical (0.6\,M$_\mathrm{\odot}$ --
log\,g=8 dex) single  white dwarfs \citep{Hollands2018, McCleery2020}.
This is  not surprising  since the  evolution of  a large  fraction of
these white  dwarfs was likely  truncated when their  progenitors were
ascending the giant  branch and the systems entered the  CE phase.  In
particular, we can  observe a considerably larger  number of extremely
low-mass  (ELM) white  dwarfs (M  $< 0.3$\,M$_\mathrm{\odot}$)  in the
\emph{Gaia}  sample.   It   is  not  clear  how   ELMs  with  low-mass
main-sequence star companions would form.  Current evolutionary models
of ELMs  in double-degenerate  systems predict  that the  formation of
ELMs with masses lower than $\simeq$0.20\,$M_{\odot}$ is likely due to
non-conservative   stable  mass   transfer  via   Roche-lobe  overflow
\citep{Sun+Arras2018, Li2019}.  This is  based on the estimated energy
that  would be  required to  expel the  envelope in  a common-envelope
scenario, which would be too large  and cause the system to eventually
merge.   However, in  our  case, the  low mass  of  the main  sequence
companion would  lead to an unstable  mass transfer.  In this  way, if
these systems  were formed  by common-envelope evolution,  somehow the
energetics should be able to eject the envelope.  Up to our knowledge,
this scenario  has not  been explored  so far.  ELM objects  have been
efficiently identified in  the SDSS as apparently  single white dwarfs
\citep{Brown2010, Brown2012,  Brown2020}.  This  is because  ELM white
dwarfs have  considerably larger radii  which, together with  the fact
that  the SDSS  seldom  targets cool  white  dwarfs for  spectroscopic
follow-up, implies  that their  companions (being these  main sequence
stars or other white dwarfs)  are generally completely out-shined.  In
the volume-limited  \emph{Gaia} sample, ELM white  dwarfs are detected
with  main  sequence  companions  only   if  these  white  dwarfs  are
sufficiently cool for not to out-shine them.  Indeed, ELM white dwarfs
in the  \emph{Gaia} sample  have effective  temperatures in  the range
3\,000-9\,500\,K, with an average of 5\,600$\pm$2\,100\,K.

Finally,  the   comparison  between   the  secondary   star  effective
temperatures   reveals  that   the  \emph{Gaia}   sample  contains   a
considerably larger  fraction of cool companions  ($\la$2800\,K). This
is a simple consequence of  the fact that \emph{Gaia} allows detecting
a larger number of cool white dwarfs in WDMS systems. These cool white
dwarfs  can only  be  observed  together with  cool  M  stars if  both
components contribute  similarly to the  SEDs, otherwise the  M dwarfs
would overwhelm  the fluxes of  the white  dwarfs. It is  important to
also note that 8 of these companions have effective temperatures below
2250\,K,  the commonly  accepted limit  which separates  main sequence
stars from  brown dwarfs  \citep{Pecaut+Mamajek2013, Kirkpatrick2021}.
Further spectroscopic  observations are required to  confirm the brown
dwarf nature of these objects.

It  has to  be emphasized  that the  stellar parameters  of the  white
dwarfs  in   \emph{Gaia}  WDMS  binaries  have   been  obtained  using
hydrogen-rich (DA) cooling sequences, an  assumption that has not been
tested  due to  the  lack  of spectroscopy.  Then,  one  may expect  a
fraction  of  non-DA white  dwarfs  in  the  sample (mainly  DBs  with
helium-rich  atmospheres, which  are the  second most  common type  of
white dwarfs after the  DAs, see e.g. \citealt{Koester2015}). However,
no single PCEB with  a DB white dwarf has been  found yet, most likely
because the  white dwarfs in  these systems accrete material  from the
wind     arising      from     their      main-sequence     companions
\citep{Parsons2013}.  This material  is  expected to  form a  hydrogen
layer on the white dwarf's atmosphere,  thus converting a DB into a DA
white dwarf.

\section{The completeness and the space density of the \emph{Gaia} unresolved WDMS binary sample}

Despite  the  fact   that  our  \emph{Gaia}  WDMS   binary  sample  is
volume-limited, it  is not  guaranteed to be  complete. First  of all,
this is  because we have focused  on identifying systems in  which the
two  stellar components  contribute in  relatively similar  amounts of
flux to  the SEDs. In second  place, as in any  other observed sample,
selection effects may induce some biases in the final sample.

In order  to estimate  the completeness  of the sample  as well  as to
derive a  space density of close  WDMS binaries, we use  a Monte Carlo
population synthesis code specifically designed to simulate the single
\citep{GB1999, GB2004, Torres2005, Jimenez2018, Torres2021} and binary
\citep{Camacho2014,  Cojocaru2017}  white  dwarf  populations  in  the
Galaxy. It is important to point  out here that this analysis shall be
understood  as  a  first  approximation   to  the  estimation  of  the
completeness of the sample. That is because the space parameter of the
white  dwarf binary  population is  not fully  determined and  several
discrepancies  arise  between  the observations  and  the  theoretical
models  (e.g. \citealt{Toonen2017}).  Consequently, it  is beyond  the
scope of the  present work to comprehensively analyze the  full set of
theoretical models. Instead,  we will focus on a  reference model that
will provide us with a first estimate.

Here we present the main ingredients of our population synthesis model
and  leave  the  details  and  the analysis  of  other  models  for  a
forthcoming  paper. The  single white  dwarf population  consist in  a
three-component Galactic model, i.e., thin and thick disk and halo, as
described in  \citet{Torres2019}. To this  population we add  a binary
population implemented through the BSE (Binary Stellar Evolution) code
of    \citet{Hurley2002}    with    some   updates    introduced    in
\citet{Camacho2014}, \citet{Cojocaru2017}  and \citet{Canals2018}. The
code incorporates  an exhaustive list of  fundamental binary evolution
ingredients,  such  as  mass   transfer,  wind  loss,  common-envelope
evolution, angular-momentum  loss, magnetic braking  and gravitational
wave radiation emission, among other  effects. A binary fraction of 50
per cent  is adopted.  The  primary masses  of the binary  systems are
randomly   chosen    following   the   initial   mass    function   of
\citet{Kroupa2001}  (also adopted  for  the  single star  population),
while   secondary   masses   follow   a  flat   distribution   as   in
\citet{Cojocaru2017}. The  orbital separations and  the eccentricities
follow  a logarithmically  flat distribution  \citep{Davis2010} and  a
linear  thermal  distribution \citep{Heggie1975},  respectively.   For
those systems that  evolve into a common envelope we  use the standard
$\alpha_{\mathrm{CE}}$-formalism \citep{Iben+Livio1993}  following the
assumptions      detailed       in      \cite{Camacho2014},      where
$\alpha_{\mathrm{CE}}$  is the  efficiency  of the  orbital energy  in
ejecting      the      envelope      (assumed     to      be      0.3;
\citealt{Zorotovic2010}). Systems  are then evolved into  present time
and their \emph{Gaia} EDR3 magnitudes are computed according to the La
Plata tracks for white  dwarfs (Section\,\ref{s-param}) and the PARSEC
tracks   for  main   sequence  stars   \citep{Bressan2012,  Chen2014}.
Photometric and astrometric errors are  added to the simulated objects
following the \emph{Gaia}'s performance.  The synthetic sample is then
normalized to $15\,753$, which is the total number of \emph{Gaia} EDR3
sources  found in  the white  dwarf loci  of the  \emph{Gaia} EDR3  HR
diagram (i.e.  the area defined  by the first condition of Equation\,2
but  extended   up  to  $G_\mathrm{BP}-   G_\mathrm{RP}\le2.0$)  after
applying the  same astrometric  and photometric selection  criteria as
that for the WDMS binary  population. Finally, we assume as unresolved
systems those for  which the angular sky separation is  smaller than 2
arcsec.   In  these  cases  the final  magnitudes  are  obtained  from
combining the individual fluxes of the two components.

It is worth mentioning that the claimed angular separation to separate
pairs   of   stars  by   \emph{Gaia}   EDR3   is  $\simeq$1.5   arcsec
\citep{Gaia2020}.  This limit  can increase  to 2  arcsec for  samples
selected with  good astrometric and  photometric parameters, as  it is
our case (Torres et al., in  preparation).  Hence the adopted value of
2 arcsec in this work.

Once the  simulation is carried  out, we have  a full modeling  of the
different sub-populations:  resolved and unresolved WDMS  binaries, as
well  as  resolved  and   unresolved  double  degenerate  systems.  In
Figure\,\ref{fig-HR-simul} we  show the  HR diagram that  results from
our simulations within  100\,pc for the sub-population  of interest in
this work, i.e. unresolved WDMS binaries.

\subsection{The completeness of the \emph{Gaia} WDMS binary sample}
\label{s-compl}

The simulations presented in the previous section allow us to estimate
the  completeness of  the  unresolved \emph{Gaia}  WDMS binary  sample
studied in this work. Inspection of Figure\,\ref{fig-HR-simul} clearly
reveals that the vast majority of WDMS binaries are entirely dominated
by the flux  of the main sequence companion, since  their positions in
the  HR diagram  are very  similar to  those of  single main  sequence
stars. As  a consequence,  only $\approx  140\pm12$ (the  error arises
after the normalization) of all the synthetic unresolved WDMS binaries
fall  within  our cuts,  that  is  $\simeq$9  per  cent of  the  total
sample. These 140 objects represent the WDMS binary population that we
should  have identified  in this  work using  the methods  outlined in
Sections\,\ref{s-sample}  and   \,\ref{s-param}.   Since   our  sample
consists of 112 WDMS binary good candidates, then the completeness can
be initially estimated as $\simeq\,80\pm9$ per cent.

Based on our  simulations we conclude that only $\simeq9$  per cent of
the  total unresolved  WDMS binary  population  in the  Galaxy can  be
efficiently identified  by our  method and  that our  catalogue (which
represents the  tip of the iceberg)  seems to be highly  complete. The
remaining $\simeq91$  per cent represent the  underlying population of
WDMS  binaries that  are heavily  dominated by  the flux  of the  main
sequence companion. Detecting such WDMS binaries is a challenging task
that   requires   UV    photometry   \citep{Maxted2009,   Parsons2016,
  Rebassa2017,    Ren2020,    Hernandez2021}    and/or    spectroscopy
\citep{Parsons2016}.

We also made use of the population synthesis simulation presented here
to estimate the  expected fraction of wide WDMS  binaries that evolved
avoiding  mass transfer  episodes but  have angular  separations of  2
arcsec or less, i.e. the limit at  which we consider the systems to be
spatially unresolved. We found that  these systems form $\simeq$28 per
cent of  the total  synthetic WDMS binary  population within  our WDMS
selected region in the HR diagram and with distances up to 100\,pc.

\subsection{The space density of close WDMS binaries}

We  have identified  112 unresolved  WDMS binary  candidates within  a
volume of 100\,pc, which translates into a space density of unresolved
(likely  PCEBs) binaries  consisting  of  a white  dwarf  plus a  main
sequence  companion  of $\simeq(2.7\pm1.6)\times  10^{-5}$\,pc$^{-3}$.
Given that  we can clearly not  rule out the possibility  that some of
our systems  are actually  wide but unresolved  binaries that  did not
evolve through CE evolution (see  previous Section), this value should
be considered as an upper limit.  \citet{Inight2021}, who analysed the
SDSS  WDMS  binary  sample  observed by  \emph{Gaia}  within  300\,pc,
derived  a  space   density  of  1.2--2.5$\times  10^{-6}$\,pc$^{-3}$,
although they claim  this result should be multiplied by  3-4 in order
to account for the SDSS selection effects.  By doing this, one obtains
a value  of 0.5--1$\times 10^{-5}$\,pc$^{-3}$,  which is lower  but of
the      same       order      to      our       estimated      value.
\citet{Schreiber+gaensicke2003} analysed a considerably smaller sample
of close  WDMS binaries and  derived a space density  of 0.6--3$\times
10^{-5}$\,pc$^{-3}$, in agreement  with our result. If  we correct our
space density estimate by the completeness of our sample ($\simeq$80),
then    this     value    increases     to    $\simeq(3.3\pm1.8)\times
10^{-5}$\,pc$^{-3}$.   If  we  further  consider that,  based  on  our
simulations,  the  sample  of  WDMS binaries  analysed  in  this  work
represents only $\simeq$9 per cent  of the underlying population, then
we estimate  an upper limit to  the total space density  of close WDMS
binaries   of   $\simeq(3.7\pm1.9)\times   10^{-4}$\,pc$^{-3}$.   This
represents  $\simeq$7  per  cent  of the  white  dwarf  space  density
calculated by \citet{Jimenez2018}.

\section{Conclusions}

Close  binaries  consisting of  a  white  dwarf  and a  main  sequence
companion, i.e.  WDMS  binaries, are important objects  to constrain a
wide variety of  open problems in modern  astronomy.  However, despite
the fact  that the  SDSS has  been very  prolific at  identifying such
systems  for follow-up  studies  in  the last  years,  this sample  is
heavily affected by selection effects. To overcome this issue, in this
work we have used the high  astrometric and photometric quality of the
recent \emph{Gaia} Early  Data Release 3 to  identify a volume-limited
sample of 112  unresolved ($\simeq$72 per cent of them  expected to be
close) WDMS binaries  within 100\,pc.  We have  measured their stellar
parameters  (effective  temperatures,   luminosities,  radii,  surface
gravities and masses)  via fitting their SEDs with  a two-body fitting
algorithm  implemented   in  VOSA.   We  find   that  the  \emph{Gaia}
volume-limited sample  consists of  intrinsically cooler  white dwarfs
and main sequence companions, some of which are in fact expected to be
brown dwarfs.   Moreover, the \emph{Gaia} sample  reveals a population
of  extremely  low-mass white  dwarfs  which  is  absent in  the  SDSS
magnitude-limited  sample. Thus,  the sample  identified here  clearly
helps  in   overcoming  the  selection  effects   affecting  the  SDSS
population.  It  has to be  emphasised however that  follow-up optical
spectroscopy is recommended  to confirm the hypothesis  stated in this
work.   Using a  Monte Carlo  population synthesis  code we  find that
$\simeq$91 per cent of the  total unresolved WDMS binary population is
expected to  be fully dominated by  the emission of the  main sequence
companions. Efficiently identifying these systems requires ultraviolet
photometry and/or spectroscopy.  We  also find strong indications that
our  catalogue, even  though it  represents  a small  fraction of  the
underlying population,  is highly  complete ($\simeq80\pm9$  per cent)
and  is not  affected by  important selection  effects.  Finally,  our
population synthesis study allows us to estimate an upper limit to the
total space density of close WDMS binaries of $\simeq(3.7\pm1.9)\times
10^{-4}$\,pc$^{-3}$.

\section*{Acknowledgements}

ARM acknowledges financial support from the MINECO under the Ram\'on y
Cajal program  (RYC-2016-20254). ST  and ARM acknowledge  support from
the  MINECO  under the  AYA2017-86274-P  grant,  and the  AGAUR  grant
SGR-661/2017.   ESM and  FJE  acknowledge financial  support from  the
MINECO under the AYA2017-86274-P  grant. FJE acknowledges support from
the H2020 ESCAPE project (Grant  Agreement no.  824064).  LMC, LGA and
AHC  acknowledge   support  from  AGENCIA  through   the  Programa  de
Modernizaci\'on Tecnol\'ogica BID 1728/OC-AR, and from CONICET through
the PIP  2017-2019 GI grant.  This  publication makes use of  VOSA and
SVO DiscTool, developed under  the Spanish Virtual Observatory project
supported from  the Spanish MINECO through  grant AyA2017-84089.  This
research has made use of Aladin sky atlas developed at CDS, Strasbourg
Observatory,     France     \citep{Bonnarel00,    Boch14}.      TOPCAT
\citep{Taylor05}  and STILTS  \citep{Taylor06} have  also been  widely
used in this paper.

We  thank the  anonymous  referee for  the  helpful suggestions.   The
authors are greatly indebted to Detlev Koester for sharing his grid of
model atmosphere white  dwarf spectra. The authors  also thank Roberto
Raddi  for  sharing  the  grid  of  white  dwarf  absolute  magnitudes
calculated for the \emph{Gaia} EDR3 bandpasses.

This work  has made use of  data from the European  Space Agency (ESA)
mission {\it  Gaia} (\url{https://www.cosmos.esa.int/gaia}), processed
by  the {\it  Gaia}  Data Processing  and  Analysis Consortium  (DPAC,
\url{https://www.cosmos.esa.int/web/gaia/dpac/consortium}).    Funding
for the DPAC has been provided by national institutions, in particular
the  institutions   participating  in  the  {\it   Gaia}  Multilateral
Agreement.

\appendix
\section{Data availability}
\label{Append}
In order to  facilitate the usage of the information  included in this
paper,   an   archive   system   that   can   be   accessed   from   a
webpage\footnote{\url{http://svo2.cab.inta-csic.es/vocats/wdw4}}    or
through               a              Virtual               Observatory
ConeSearch\footnote{e.g. \url{http://svo2.cab.inta-csic.es/vocats/wdw4/cs.php?RA=157.163\&DEC=73.180\&SR=0.1}}
has been built.

The  archive system  implements a  very simple  search interface  that
permits queries  by coordinates  and range of  effective temperatures,
surface gravities, luminosities,  radii and masses. The  user can also
select the maximum number of sources to return (with values from 10 to
unlimited).  The  result of  the query  is a HTML  table with  all the
sources  found in  the  archive fulfilling  the  search criteria.  The
result can also be downloaded as a VOTable or a CSV file.

Detailed information on the output  fields can be obtained placing the
mouse over the question mark (``?'')  located close to the name of the
column.        The       archive       also       implements       the
SAMP\footnote{\url{http://www.ivoa.net/documents/SAMP/}}       (Simple
Application  Messaging)  Virtual  Observatory protocol.   SAMP  allows
Virtual Observatory applications  to communicate with each  other in a
seamless and transparent manner for the user. This way, the results of
a query can  be easily transferred to other VO  applications, such as,
for instance, Topcat.


\typeout{}








\bsp	
\label{lastpage}
\end{document}